# Magnetism and local structure in low-dimensional, Mott insulating GdTiO$_3$

Jack Y. Zhang*, Clayton A. Jackson*, Santosh Raghavan, Jinwoo Hwang, and Susanne Stemmer

*Materials Department, University of California, Santa Barbara, CA 93106-5050, USA*

*These authors contributed equally to this work

**Abstract**

Cation displacements, oxygen octahedral tilts, and magnetism of epitaxial, ferrimagnetic, insulating $GdTiO_3$ films sandwiched between cubic $SrTiO_3$ layers are studied using scanning transmission electron microscopy and magnetization measurements. With decreasing $GdTiO_3$ film thickness, structural ($GdFeO_3$-type) distortions are reduced, concomitant with a reduction in the Curie temperature. Ferromagnetism persists to smaller deviations from the cubic perovskite structure than is the case for the bulk rare earth titanates. The results indicate that the FM ground state is controlled by the narrow bandwidth, exchange and orbital ordering, and only to second order depends on amount of the $GdFeO_3$-type distortion.

Perovskite rare-earth titanates are key materials to understand emergent phenomena caused by the coupling of the electron, lattice, spin, and orbital degrees of freedom. They are strongly correlated Mott insulators, with a single electron occupying the Ti $t_{2g}$ orbitals. Magnetic ordering is closely coupled with distortions and tilts of the Ti-O octahedra in the orthorhombic $GdFeO_3$ structure (space group Pbnm) that all rare earth titanates adopt, and which removes the orbital degeneracy [1-3]. Two distinct types of orbital polarization, namely ferro-orbital and antiferro-orbital ordering, have been reported, and are compatible with Pbnm symmetry [4-9]. Ferro-orbital ordering is found in the antiferromagnetic (AFM) titanates, which also exhibit smaller $GdFeO_3$-type distortions (A = La…Sm in the chemical formula $ATiO_3$), whereas antiferro-orbital ordering is found in the ferromagnetic (FM) titanates that also have larger distortions (A = Gd…Y).

The $GdFeO_3$ structure is characterized by $a^-a^-b^+$ type octahedral tilts in Glazer notation [10]. The two degrees of freedom in the Pbnm space group ($x$ and $y$) allow the A-site cations to shift to a more energetically favorable position. The amount of displacement depends on the octahedral geometry [11]. The degree of the $GdFeO_3$-type distortion appears to be a primary factor determining the transition from AFM to FM ordering, and the ordering temperature [1, 3, 12, 13]. Nevertheless, the relative roles of orbital-lattice coupling and structural distortions in this transition remain a subject of significant debate [6, 12, 14, 15].

Epitaxial mismatch strains in heterostructures and the need for interfacial connectivity of the oxygen octahedra offer distinct and precise ways of tuning octahedral rotations and distortions without chemical substitution [16, 17]. This may allow for controlling orbital-lattice coupling, and thus the magnetic properties, as well as insights into materials physics not possible with bulk materials.

In this Rapid Communication, we report on the local structure and magnetism of $GdTiO_3$ films that are sandwiched between cubic $SrTiO_3$ with the goal to understand how rigidly the magnetic interactions are coupled to the octahedral tilts in a prototype FM rare earth titanate. In $GdTiO_3$, the ferromagnetic Ti array couples antiferromagnetically to the Gd ions, resulting in net ferrimagnetism [12, 13, 18]. The AFM Gd-O-Ti interactions are believed to be weaker than the ferromagnetic Ti-O-Ti interactions [12]. $GdTiO_3$ is just on the FM side of the FM-AFM phase boundary; therefore, if the FM character is sensitive to octahedral rotations and distortions, significant effects on its magnetism with structural modifications may be expected.

In a previous study we have shown that approximately 1-3 GdO planes near the interface with $SrTiO_3$ exhibit a significant reduction in Gd displacements, and thus octahedral tilts, while in the interior of the $GdTiO_3$ layers these displacements agreed reasonably well with bulk values [19]. Thus the need to maintain interfacial oxygen octahedral connectivity is mostly accommodated within the interfacial $GdTiO_3$. This suggests that by further decreasing the thickness of the $GdTiO_3$, the octahedral distortions in the entire $GdTiO_3$ layer can be modified.

$GdTiO_3$ films and $GdTiO_3/SrTiO_3$ superlattices were grown on (001) $(La_{0.3}Sr_{0.7})(Al_{0.65}Ta_{0.35})O_3$ (LSAT) by hybrid molecular beam epitaxy (MBE) [20, 21]. A 20 nm $GdTiO_3$ film was grown directly on LSAT, while $GdTiO_3$ layers of 3.5 nm, 2.4 nm, and 2.0 nm thickness (10, 7, and 6 GdO layers, respectively) were grown in a superlattice structure with 5 nm of $SrTiO_3$ spacers. Superlattices contained either 5 or 10

GdTiO$_3$ layers, and thus approximately the same amount of GdTiO$_3$, by volume, as the 20 nm sample. They had 10 nm SrTiO$_3$ buffers and caps, respectively.

The magnetization was measured in a SQUID magnetometer (Quantum Design) with the magnetic field in the plane of the film. Cross-section transmission electron microscopy (TEM) foils were prepared by focused ion beam (5 kV Ga ions) and imaged using a field emission FEI Titan S/TEM with a super-twin lens (C$_s$ = 1.2 mm) at 300 kV, using a 1024×1024 frame size and 30 μs dwell time. The convergence angle was 9.6 mrad. A deviation angle, (180° − $\theta$), was used to quantify the A-site (Gd, Sr) displacements, where $\theta$ is the angle between three successive A-site cations [see Fig. 1(a)]. A-site positions were determined from multiple high angle annular dark field (HAADF) images of each layer. Atomic centroid positions were extracted using a custom MATLAB algorithm [22]. Orientation domains in GdTiO$_3$ are present (see refs. [19, 20]). All images were taken along [110]$_O$, as A-site displacements can be discerned along this direction [19]. While MBE provides near-monolayer thickness control, substrate miscut and surface steps cause uncertainties of ± 1 atomic plane in estimates of the layer thickness along the growth direction. The thicknesses given here represent the average number of atomic planes in TEM. Only images of layers with the nominal thicknesses were selected for further analysis. Representative HAADF images of different GdTiO$_3$ thicknesses are shown in Fig. 1. Octahedral tilts were characterized using position averaged convergent beam electron diffraction (PACBED) [17, 23, 24]. PACBED patterns were recorded from areas of ~ four pseudocubic unit cells (slightly larger than the primitive orthorhombic unit cell projection), at the center of each GdTiO$_3$ film.

A-site cation displacements for each AO plane along the growth direction are shown in Fig. 2 for different GdTiO$_3$ thicknesses. Shaded areas indicate the GdTiO$_3$ layer, which can easily be identified from HAADF image intensities. The dashed line represents the average (~100 atomic rows over four images) deviation angle, ~15°, of the 20 nm GdTiO$_3$ film, which matches that of bulk GdTiO$_3$. As discussed elsewhere [19], SrO planes show no Sr displacements (the apparent deviation angle of ~1° is due to noise and instability, and serves as a measure of the error). About 1-3 GdO planes near the interface show reduced deviation angles in all samples, as discussed above. GdTiO$_3$ quantum wells of 3.5, 2.4, and 2.0 nm (10, 7, and 6 GdO layers, respectively) show reduced deviation angles also *in the films' interior*, not just at the interface. For the 3.5 and 2.4 nm films, the deviation angle is constant at the center, slightly reduced from bulk for the 3.5 nm film (~14°), and with a significantly reduced value for the 2.4 nm film (~11°). In the 2.0 nm film, the deviation angle is ~10° in the center and then continuously decreases towards the interface.

To confirm that Gd displacements correlate with the octahedral tilts and distortions, as they do in bulk, PACBED was carried out. Figure 3 shows simulated [110]$_O$ PACBED patterns (Kirkland multislice code [25]) for different octahedral tilts (rows) and Gd displacements (columns). It can be seen that the symmetry and features in PACBED are sensitive to the octahedral tilts, whereas the effects of Gd displacements are minor. Patterns without octahedral tilts (top row) show a dark concave octagonal shape, along with four "cross-shaped" regions within the central disk, while tilted patterns (bottom row) appear more square-like, "lens-like" in the center and triangular corners. These features remain consistent regardless of Gd displacements. Figure 4 compares experimental and simulated PACBED patterns. For the simulations, the octahedral tilts and distortions are varied; the degree of distortion is based on the Gd-site displacements obtained from Fig. 2, and interpolation using bulk rare-earth data [12]. The top halves of the simulated patterns are convolved with a Gaussian function to account for the experimental point spread function and show good agreement with the experiment. As the octahedral rotations decrease, the $GdTiO_3$ PACBED becomes more "$SrTiO_3$-like," (i.e., cubic).

Figure 5(a) shows the magnetization of each sample as a function of temperature under a constant field of 100 Oe. The magnetization hysteresis at 2 K is shown in Fig. 5(b). The measured magnetization includes the diamagnetic and paramagnetic responses from the $SrTiO_3$ layers, the LSAT and a Ta backing layer. Although all samples were similar in size and contained comparable amounts of $GdTiO_3$, small size and thickness variations of substrate and backing layer are unavoidable. Isolating the $GdTiO_3$ response could not done due to the superlattice structure. Therefore, conclusions about parameters that depend on the volume (saturation magnetization) should be made with care. The Curie temperature ($T_c$) and coercivity are, however, properties of only the FM $GdTiO_3$. The $T_c$ of the 20 nm film (~ 30 K) agrees well with the bulk [12, 13, 18]. All $GdTiO_3$ layers with thicknesses greater than 2.0 nm are FM, but their $T_c$ decreases continuously with decreasing thickness (see arrows).

Comparing Figs. 2 and 5 reveals that the reduction in octahedral tilts causes a decrease in $T_c$. This behavior is expected by analogy with the bulk rare earth titanates, which show a decrease in $T_c$ with increasing bandwidth (reduced distortions), consistent with band ferromagnetism [26]. However, a quantitative comparison of film and bulk data reveals significant differences. Figure 6 shows the magnetic phase diagram as a function of deviation angle for thin films and bulk, respectively. In bulk, a transition from AFM to FM ordering occurs between Gd and Sm, at a deviation angle of ~15°. In contrast, in the layers, the critical angle for FM behavior to vanish is 10.5 ± 1°. This deviation angle is comparable to that of AFM $LaTiO_3$, which has the smallest $GdFeO_3$-type distortion among all the rare earth titanates, and is barely insulating.

The important conclusion is that FM ordering/anti-ferro-orbital ordering are not as strongly dependent on the orthorhombic distortion as may be implied from (naïve) interpretation of the bulk phase diagram. Rather, the results support a picture of a direct interaction between orbital ordering, which determines the magnetism, and the lattice, somewhat independent from the degree of orthorhombic distortion. This interpretation is in agreement with recent reports of structural anomalies at the magnetic ordering temperature, which also support a direct lattice-orbital coupling [12]. Takubo and co-authors found that in the AFM rare earth titanates, the orbital ordering changes at the ordering temperature, also suggesting a direct interaction [6]. The FM rare earth titanates thus appear to closely match models of narrow band, insulating, one-electron systems [27-29] with a FM ground state. In these systems, antiferro-orbital ordering in conjunction with intra-atomic exchange results in ferromagnetism at a temperature below the orbital ordering temperature. The $4f^7$ configuration of the Gd ions ensures no orbital angular momentum contributions from the Gd, suggesting the interatomic exchange field, even at lower Ti-O-Ti bond angles, favors the FM ground state.

The FM ground state only vanishes in the 2.0 nm film that does not contain any continuous planes with the same Gd displacements (octahedral tilts) anymore. This may make long-range, coherent orbital ordering [14] difficult, similar to what is observed in alloys such as $La_{1-x}Y_xTiO_3$ [12] or $Sm_{1-x}Gd_xTiO_3$ [26], and may explain the vanishing $T_c$.

An open question that could not be answered is why deviation angles are already reduced in the interior of films that are still thicker than (twice) the thickness needed to accommodate the oxygen octahedral connectivity at the interface with $SrTiO_3$. Theoretical simulations that consider long-range structural coherencies, as well as possible coupling with or between the high-density two-dimensional electron gases that are located at the interfaces in the $SrTiO_3$ [30] may be needed to understand this.

Finally, we note that the results are consistent with the hypothesis that "magnetic deadlayers", widely reported for many perovskite films, are caused by interfacial structural distortions due to oxygen octahedral connectivity constraints. Suitably designed heterostructures (i.e., interfaces with a smaller degrees of tilt mismatch) may be able to mitigate this.

The authors thank Leon Balents for many helpful discussions. The microscopy studies were supported by the DOE (grant no. DEFG02-02ER45994). The magnetism studies were supported by a MURI program of the Army Research Office (Grant No. W911-NF-09-1-0398). J.Y.Z. received support from the Department of Defense through an NDSEG fellowship, and C. A. J. from the National Science Foundation through a Graduate Research Fellowship. Acquisition of the oxide MBE system used in this study was made possible through an NSF MRI grant (Award No. DMR 1126455). This work made use of facilities from the Center for Scientific Computing at the California

Nanosystems Institute (NSF CNS-0960316) and the UCSB Materials Research Laboratory, an NSF-funded MRSEC (DMR-1121053).

**References**

[1]  M. Mochizuki, and M. Imada, New J. Phys. **6**, 154 (2004).

[2]  J. G. Cheng, Y. Sui, J. S. Zhou, J. B. Goodenough, and W. H. Su, Phys. Rev. Lett. **101**, 087205 (2008).

[3]  J. B. Goodenough, and J. S. Zhou, J. Mater. Chem. **17**, 2394 (2007).

[4]  J. Akimitsu, H. Ichikawa, N. Eguchi, T. Miyano, M. Nishi, and K. Kakurai, J. Phys. Soc. Jpn. **70**, 3475 (2001).

[5]  M. Itoh, M. Tsuchiya, H. Tanaka, and K. Motoya, J. Phys. Soc. Jpn. **68**, 2783 (1999).

[6]  K. Takubo, M. Shimuta, J. E. Kim, K. Kato, M. Takata, and T. Katsufuji, Phys. Rev. B **82**, 020401 (2010).

[7]  H. Nakao *et al.*, Phys. Rev. B **66**, 184419 (2002).

[8]  R. Schmitz, O. Entin-Wohlman, A. Aharony, A. B. Harris, and E. Muller-Hartmann, Phys. Rev. B **71**, 144412 (2005).

[9]  E. Pavarini, S. Biermann, A. Poteryaev, A. I. Lichtenstein, A. Georges, and O. K. Andersen, Phys. Rev. Lett. **92**, 176403 (2004).

[10] A. M. Glazer, Acta Cryst. B **28**, 3384 (1972).

[11] P. M. Woodward, Acta Crystallogr. Sect. B **53**, 44 (1997).

[12] A. C. Komarek, H. Roth, M. Cwik, W. D. Stein, J. Baier, M. Kriener, F. Bouree, T. Lorenz, and M. Braden, Phys. Rev. B **75**, 224402 (2007).

[13] H. D. Zhou, and J. B. Goodenough, J. Phys.: Condens. Matter **17**, 7395 (2005).

[14] M. Kubota, H. Nakao, Y. Murakami, Y. Taguchi, M. Iwama, and Y. Tokura, Phys. Rev. B **70**, 245125 (2004).

[15] E. Pavarini, A. Yamasaki, J. Nuss, and O. K. Andersen, New J. Phys. **7** (2005).

[16] J. M. Rondinelli, S. J. May, and J. W. Freeland, MRS Bull. **37**, 261 (2012).

[17] J. Hwang, J. Son, J. Y. Zhang, A. Janotti, C. G. Van de Walle, and S. Stemmer, Phys. Rev. B **87**, 060101 (2013).

[18] C. W. Turner, and J. E. Greedan, J. Solid State Chem. **34**, 207 (1980).

[19] J. Y. Zhang, J. Hwang, S. Raghavan, and S. Stemmer, Phys. Rev. Lett. **110**, 256401 (2013).

[20] P. Moetakef, J. Y. Zhang, S. Raghavan, A. P. Kajdos, and S. Stemmer, J. Vac. Sci. Technol. A **31**, 041503 (2013).

[21] B. Jalan, R. Engel-Herbert, N. J. Wright, and S. Stemmer, J. Vac. Sci. Technol. A **27**, 461 (2009).

[22] J. M. LeBeau, and S. Stemmer, Ultramicroscopy **108**, 1653 (2008).

[23] J. M. LeBeau, S. D. Findlay, L. J. Allen, and S. Stemmer, Ultramicroscopy **110**, 118 (2010).

[24] J. Hwang, J. Y. Zhang, J. Son, and S. Stemmer, Appl. Phys. Lett. **100**, 191909 (2012).

[25] E. J. Kirkland, *Advanced Computing in Electron Microscopy* (Springer, New York, 2010).

[26] G. Amow, J. S. Zhou, and J. B. Goodenough, J. Solid State Chem. **154**, 619 (2000).

[27] L. M. Roth, Phys. Rev. **149**, 306 (1966).

[28] S. Inagaki, J. Phys. Soc. Jpn. **39**, 596 (1975).

[29] D. I. Khomskii, and K. I. Kugel, Solid State Commun. **13**, 763 (1973).

[30] P. Moetakef *et al.*, Appl. Phys. Lett. **99**, 232116 (2011).

[31] R. D. Shannon, Acta Crystallogr. Sect. A **32**, 751 (1976).

**Figure Captions**

**Figure 1 (Color online):** (a) HAADF-STEM image of a 20 nm GdTiO$_3$ film showing Gd displacements. The angle $\theta$ is measured between three successive Gd columns. A schematic of the unit cell is superimposed. (b)-(d) Representative images of 3.5, 2.4, and 2.0 nm thick films.

**Figure 2 (Color online):** Deviation angles for each AO plane across SrTiO$_3$/GdTiO$_3$/SrTiO$_3$ interfaces with different GdTiO$_3$ thicknesses. The angle for the 20 nm film is indicated by the dashed line and is an average over ~100 GdO planes. Shaded regions indicate the extent of the GdTiO$_3$ film for each sample, determined from the HAADF image intensities.

**Figure 3 (Color online):** Simulated "GdTiO$_3$" PACBED patterns for different Gd displacements and octahedral tilts (TEM foil thickness: 18.8 nm). The numbers indicate the degree of distortion, with 0 signifying no distortion, 1 the distortion in GdTiO$_3$, and ½ corresponding to the intermediate distortion. The top-left panel corresponds to the cubic structure, while the bottom-right panel is bulk GdTiO$_3$.

**Figure 4 (Color online):** Experimental (top row) and simulated (bottom row) PACBED patterns of GdTiO$_3$ and SrTiO$_3$. White labels indicate the TEM foil thickness, and black ones the GdTiO$_3$ layer thicknesses from which the experimental data was acquired. Simulated patterns use the expected octahedral tilts from measured deviation angles. Gd

displacements were taken to be bulk-like for both GdTiO$_3$ simulations. The top half of simulated patterns include Gaussian convolution to account for detector point spread function, and show a better match to experimental patterns.

**Figure 5 (Color online):** (a) Magnetization as a function of temperature for samples with GdTiO$_3$ films of various thicknesses recorded on cooling under a field of 100 Oe. The arrows indicate $T_c$. The data from the 19 nm sample is from ref. [20]. (b) Magnetization as a function of magnetic field at 2 K.

**Figure 6 (Color online):** Measured deviation angles (open diamonds, top graph) for GdTiO$_3$ films with different film thicknesses. The FM stability region is indicated. The angles are an average of the center regions in the 3.5 and 2.4 nm quantum wells, and the peak value for 2.0 nm quantum well. The arrow represents the estimated uncertainty of ± 1°, estimated from the SrTiO$_3$ deviation angle measurements, shown only on 20 nm film data for clarity, but applies to all measurements. The bottom graph (open circles) shows the deviation angles for bulk rare-earth titanates [12] with different rare earth ionic radii [31]. Filled triangles estimate the effects of coherent substrate strain and microscope scan asymmetry (~2% difference between x and y directions, measured from cubic samples). Both change the measured lattice parameters and, hence, deviation angles. The FM and AFM stability regions are indicated.

**Figure 1**

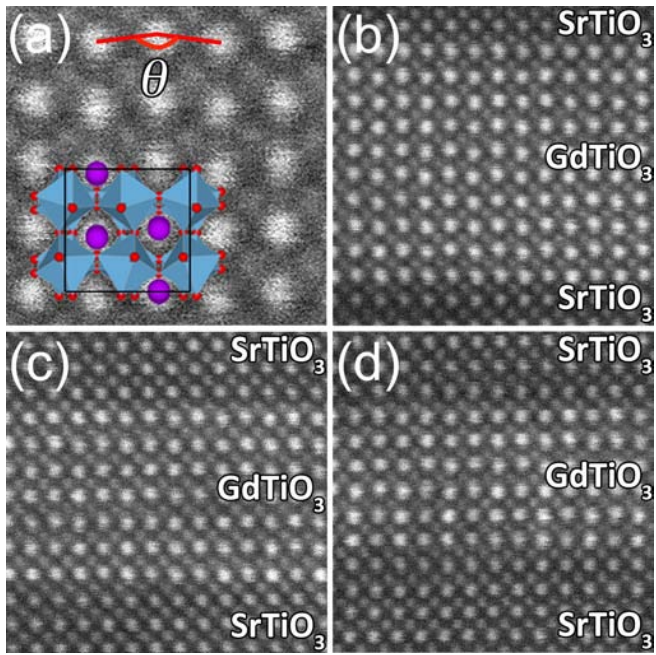

**Figure 2**

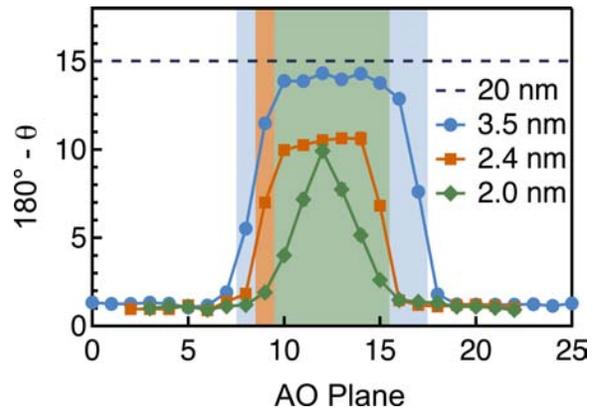

**Figure 3**

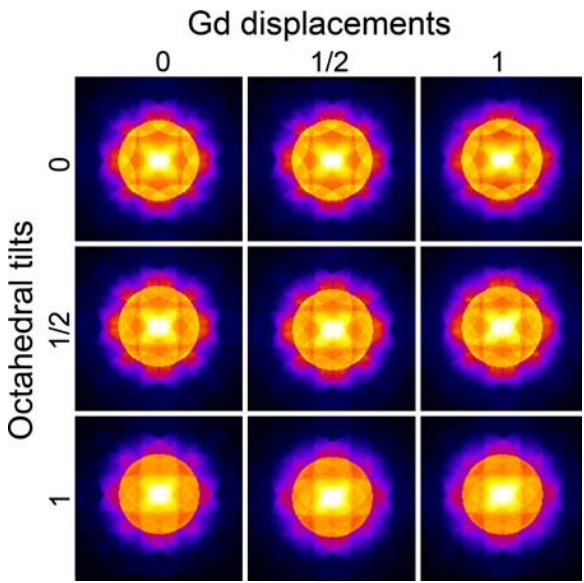

**Figure 4**

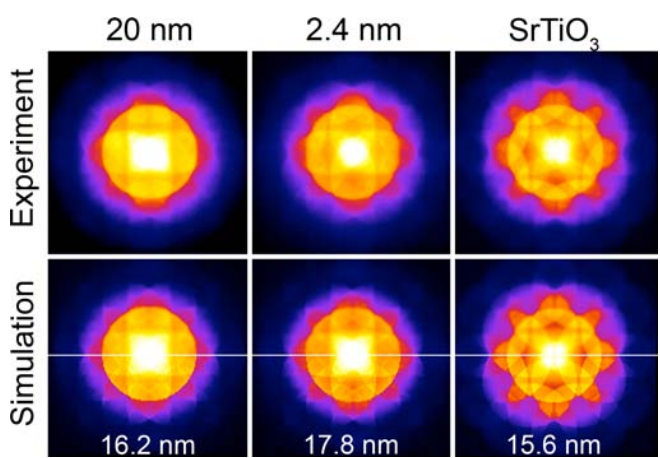

**Figure 5**

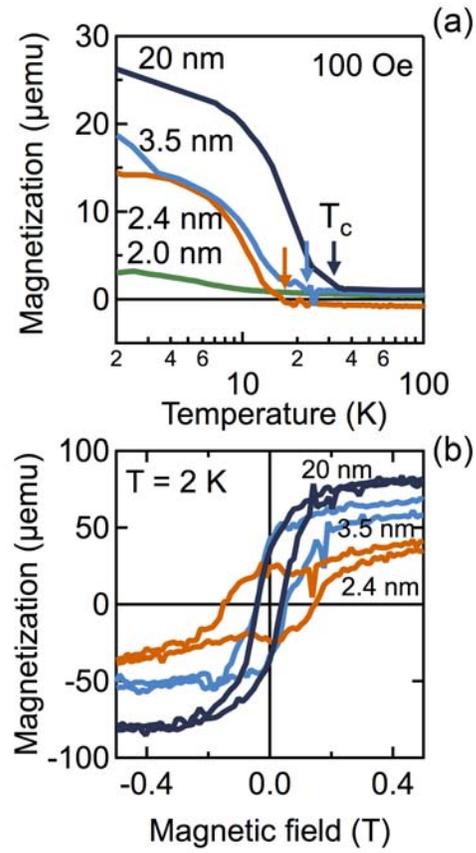

**Figure 6**

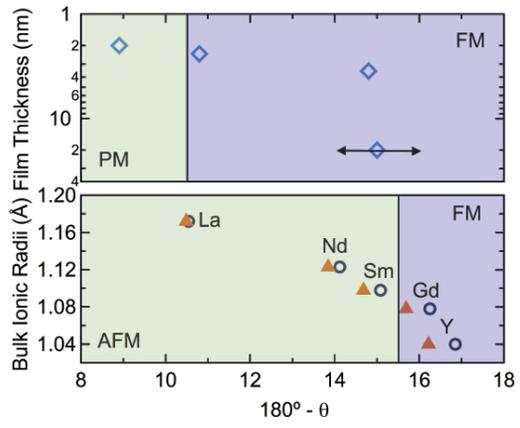